# Rigid Foldability and Mountain-Valley Crease Assignments of Square-Twist Origami Pattern

Huijuan Feng [a,b,c], Rui Peng [a,b], Shixi Zang [a,b], Jiayao Ma [a,b], Yan Chen [a,b,*]

[a] Key Laboratory of Mechanism Theory and Equipment Design of Ministry of Education, Tianjin University, 135 Yaguan Road, Tianjin, 300350, China

[b] School of Mechanical Engineering, Tianjin University, 135 Yaguan Road, Tianjin, 300350, China

[c] Reconfigurable Robotics Lab, École Polytechnique Fédérale de Lausanne (EPFL), 1015 Lausanne, Switzerland

*Corresponding author. Email: yan_chen@tju.edu.cn (Y. Chen)

**Abstract:**

Rigid foldability allows an origami pattern to fold about crease lines without twisting or stretching component panels. It enables folding of rigid materials, facilitating the design of foldable structures. Recent study shows that rigid foldability is affected by the mountain-valley crease (M-V) assignment of an origami pattern. In this paper, we investigate the rigid foldability of the square-twist origami pattern with diverse M-V assignments by a kinematic method based on the motion transmission path. Four types of square-twist origami patterns are analyzed, among which two are found rigidly foldable, while the other two are not. The explicit kinematic equations of the rigid cases are derived based on the kinematic equivalence between the rigid origami pattern and the closed-loop network of spherical $4R$ linkages. We also propose a crease-addition method to convert the rigid foldability of the non-rigid patterns. The motion compatibility conditions of the modified patterns are checked, which verify the rigid foldability of the modified patterns. The kinematic analysis reveals the bifurcation behaviour of the modified patterns. This work not only helps to deepen our understanding on the rigid foldability of origami patterns and its relationship with the M-V assignments, but also provides us an effective way to create more rigidly foldable origami patterns from non rigid ones.

# 1. Introduction

Origami, originating from the art of paper folding, has drawn increasing attention of mathematicians, scientists and engineers since the mid-1970s [1]. It has recently seen surge in a variety of engineering fields ranging from aerospace engineering [2], civil engineering [3], biomedical engineering [4] to robotics [5], where its movement and mechanical properties are essential. Considering the stiffness of constituting materials in these origami structures, rigid origami that permits continuous motion along the pre-determined creases without stretching or bending of the facets appears as an effective way for motion analysis. Rigid foldability measures the capacity of an origami pattern to fold about crease lines without twisting or stretching component facets. Several works have been done to judge the rigid foldability of origami patterns. For example, Watanabe and Kawaguchi proposed diagram and numerical methods to check the rigid foldability of several known origami patterns [6]. Tachi used some numerical algorithms to find out a family of rigidly foldable origami with quadrilateral mesh [7]. Cai et al. developed a new method to check the rigid foldability of cylindrical foldable structures by combining the quaternion rotation sequence method and the dual quaternion method [8][9]. Dai and Jones firstly modelled the paper folding by treating the creases as rotation joints and the facets as links [10]-[12], which opens up the way to analyze origami patterns from the mechanism perspective. The rigid origami around each vertex is treated as a spherical linkage in which the axes of all joints meet at a point [13][14]. An origami pattern with multiple vertices is then regarded as an assembly network of spherical linkages. Therefore, the rigid foldability of origami patterns can be analyzed based on spherical linkages. Hull adopted the spherical trigonometry to judge the rigid foldability of some origami patterns with four-crease vertices [15]. Wu and You established the rotating vector model based on the origami-spherical linkage analogy and employed quaternion and dual quaternion to study the rigid foldability of both single-vertex and multi-vertex origami patterns [16]. Streinu and Whiteley proved the rigid foldability of some single-vertex origami by linking it to spherical polygonal linkages [17]. Xi and Lien dealt with the foldability problem of origami patterns through a randomized method by modelling rigid origami as a kinematic system with closure constraints [18]. Wang and Chen modelled several origami patterns with equilateral trapezoids, general trapezoids and general quadrilaterals as spherical linkage assemblies in the closed patterned cylinder design [19]. Moreover, the general condition for rigidly foldable prismatic structures was figured out by solving the kinematics and compatibility of the mobile assemblies of spherical 4R linkages [20][21].

Recently, it has been found that the rigid foldability of origami patterns is affected by the mountain-valley crease (M-V) assignments [22]-[24]. Hull examined the problem of counting the number of valid M-V assignments for a given crease pattern and developed recursive functions for single-vertex crease patterns [15]. For multi-vertex crease patterns, Evans et al. discussed the effect of M-V assignments on the rigid foldability for several origami twists including triangle twists, quadrilateral twists and regular polygon twists [22]. Peng et al. sought the rigid foldability of double corrugated pattern with diverse M-V assignments using a general kinematic method based on the motion transmission path [23]. Feng et al. figured out all M-V assignments of the generalized triangle-twist origami pattern and analyzed their rigid foldability [24].

As one of the traditional origami patterns, the square-twist pattern has aroused substantial research interest. Kawasaki and Yoshida first proposed the square-twist pattern as an origami art and demonstrated two tessellating methods of this pattern [25]. Hull studied the square-twist origami pattern using the flat-foldable theory and found 16 different M-V assignments [15][26]. Excluding the duplicated cases, only six unique types of M-V assignments of the square-twist origami patterns exist, among which four types are well known, while the remaining two are the mirror image cases [27]. The rigid foldability of the four square-twist origami patterns was studied by introducing the fold-angle multipliers [22]. These square-twist patterns with different M-V assignments were used to generate mechanical metamaterial with negative Poisson's ratio [28]. Introducing the facet-bending property in the mechanical analysis, Silverberg et al. found that one of the square-twist patterns was bi-stable with a hidden degree of freedom [29]. The same square-twist pattern was investigated by Al-Mulla and Buehler to validate the method of replacing the facet bending by additional crease rotation [30]. Even though the square-twist origami patterns are widely used and studied, there is seldom kinematic analysis, which is essential for the performance analysis of structures or metamaterials constructed by these patterns.

In this paper, we will analyze the rigid foldability and kinematics of the square-twist origami patterns with different M-V assignments based on the kinematic motion transmission path. The explicit kinematic equations will be derived for the rigidly foldable cases. We will also propose a method to convert the non rigid patterns to rigid ones. The layout of this paper is as follows. Section 2 gives the rigid foldability analysis of square-twist origami patterns with different M-V assignments. The non-rigid square-twist patterns are converted to rigid ones by adding creases into the patterns in Sec. 3, where the rigid foldability of the modified patterns is verified. The kinematics and bifurcation behaviour analysis of the modified patterns are conducted in Sec. 4, followed by the conclusions and discussion in Sec. 5.

## 2. Square-twist patterns and their rigid foldability

In the origami art, the square-twist pattern is formed by a central square with four parallel crease-pairs radiating from each side of the central square [22]. It is composed of a twisted square and four alternatively-placed pairs of rectangles and trapezoids, as shown in Fig. 1(a). It contains four identical 4-crease vertices, noted as A, B, C and D. $a_i$, $b_i$, $c_i$ and $d_i$ ($i$=1, 2, 3, 4) represent the creases around the corresponding vertex. The sector angles of each vertex are $\alpha$, $\beta$, $\delta$ and $\gamma$. In rigid origami, the vertex can be modeled as a spherical 4R linkage as presented in Fig. 1(b). Hence, the square-twist origami pattern is kinematically equivalent to the closed-loop network of four spherical 4R linkages in Fig. 1(c). Here, links $a_1a_2$ and $a_2a_3$ in linkage A are rigidly connected to links $b_1b_2$ and $b_1b_4$ in linkage B, respectively. The similar connection method is applied to other adjacent links, such as links $b_1b_2$ & $c_1c_2$, $b_2b_3$ & $c_1c_4$, $c_1c_2$ & $d_1d_2$, $c_2c_3$ & $d_1d_4$, $d_1d_2$ & $a_1a_2$, and $d_2d_3$ & $a_1a_4$. In this manner, joints $a_2$ and $b_1$ are coaxial and have identical rotation, as well as joints $b_2$&$c_1$, $c_2$&$d_1$, and $d_2$&$a_1$.

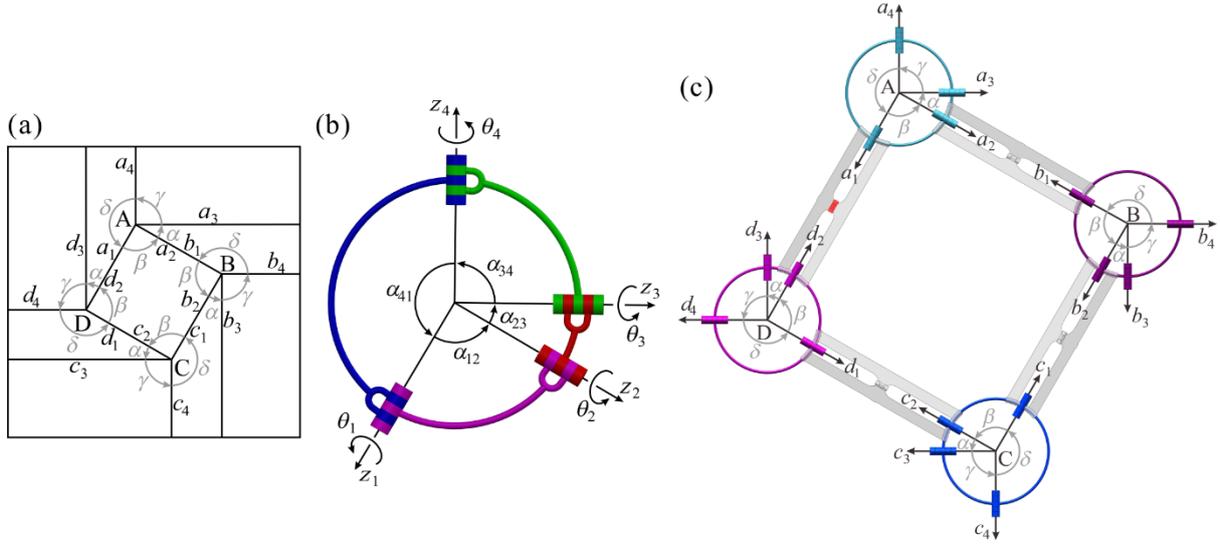

Fig. 1 Square-twist origami pattern and its linkage counterpart: (a) the square-twist origami pattern, (b) the equivalent spherical 4R linkage of the 4-crease vertex, and (c) the equivalent closed-loop network of four spherical 4R linkages for the square-twist pattern.

The spherical linkage is formed by a loop of adjacent rigid links connected by only revolute joints, where the lengths and offsets of all links are zero. Using the Denavit-Hartenberg (D-H) notation [31] and the matrix method for kinematic analysis, the closure equation for the spherical linkage with $n$ links is

$$\mathbf{Q}_{21} \cdot \mathbf{Q}_{32} \cdots \mathbf{Q}_{1n} = \mathbf{I}_3, \qquad (1)$$

where the transformation matrix $\mathbf{Q}_{(i+1)i}$ that transforms the expression in the $(i+1)^{\text{th}}$ coordinate system to the $i^{\text{th}}$ coordinate system is

$$\mathbf{Q}_{(i+1)i} = \begin{bmatrix} \cos\theta_i & -\cos\alpha_{i(i+1)} \cdot \sin\theta_i & \sin\alpha_{i(i+1)} \cdot \sin\theta_i \\ \sin\theta_i & \cos\alpha_{i(i+1)} \cdot \cos\theta_i & -\sin\alpha_{i(i+1)} \cdot \cos\theta_i \\ 0 & \sin\alpha_{i(i+1)} & \cos\alpha_{i(i+1)} \end{bmatrix}, \qquad (2)$$

and when $i+1 > n$, it is replaced by 1. Here, $\alpha_{i(i+1)}$ is the sector angle between the $i^{\text{th}}$ and $(i+1)^{\text{th}}$ joints, and $\theta_i$ is the rotation angle of the $i^{\text{th}}$ joint. For the spherical 4R linkage that satisfies the flat-foldable condition ($\alpha_{12} = \pi - \alpha_{34}$, $\alpha_{23} = \pi - \alpha_{41}$) [32] in Fig. 1(b), the relationships between the kinematic variables $\theta_i$ and $\theta_{i+1}$ ($i=1, 2, 3, 4$) can be obtained by solving Eq. (1) as follows.

$$\cos\alpha_{(i+1)(i+2)} \cdot \sin\alpha_{(i-1)i} \cdot \sin\alpha_{i(i+1)} \cdot \cos\theta_i + \cos\alpha_{(i-1)i} \cdot \sin\alpha_{i(i+1)} \cdot \sin\alpha_{(i+1)(i+2)} \cdot \cos\theta_{i+1}$$
$$+\cos\alpha_{i(i+1)} \cdot \sin\alpha_{(i+1)(i+2)} \cdot \sin\alpha_{(i-1)i} \cdot \cos\theta_i \cdot \cos\theta_{i+1} - \sin\alpha_{(i+1)(i+2)} \cdot \sin\alpha_{(i-1)i} \cdot \sin\theta_i \cdot \sin\theta_{i+1} \quad (3)$$
$$+\cos\alpha_{(i+2)(i+3)} - \cos\alpha_{i(i+1)} \cdot \cos\alpha_{(i+1)(i+2)} \cdot \cos\alpha_{(i-1)i} = 0$$

In general, $-\pi \leq \theta_i \leq \pi$. Yet physically, the paper cannot penetrate through each other. So we have $0 \leq \theta_i \leq \pi$ for the mountain crease, and $-\pi \leq \theta_i \leq 0$ for the valley one.

For the 4-crease vertex in the square-twist pattern, the geometrical parameters of the equivalent spherical 4R linkage are defined as

$$\alpha_{12} = \beta = \frac{\pi}{2}, \alpha_{23} = \alpha, \alpha_{34} = \gamma = \frac{\pi}{2}, \alpha_{41} = \delta = \pi - \alpha. \quad (4)$$

Here, $\alpha$ is the twist angle and we have $\alpha \in (0, \pi/2)$. Substituting Eq. (4) to Eq. (3), the solution of the kinematic relationship is expressed as

$$\tan\frac{\theta_2}{2} = \frac{\cos\alpha}{1-\sin\alpha}\tan\frac{\theta_1}{2}, \tan\frac{\theta_3}{2} = \frac{-\cos\alpha}{1+\sin\alpha}\tan\frac{\theta_2}{2},$$
$$\tan\frac{\theta_4}{2} = \frac{-\cos\alpha}{1-\sin\alpha}\tan\frac{\theta_3}{2}, \tan\frac{\theta_1}{2} = \frac{\cos\alpha}{1+\sin\alpha}\tan\frac{\theta_4}{2}. \quad (5a)$$

$$\tan\frac{\theta_2}{2} = \frac{-\cos\alpha}{1+\sin\alpha}\tan\frac{\theta_1}{2}, \tan\frac{\theta_3}{2} = \frac{-\cos\alpha}{1-\sin\alpha}\tan\frac{\theta_2}{2},$$
$$\tan\frac{\theta_4}{2} = \frac{\cos\alpha}{1+\sin\alpha}\tan\frac{\theta_3}{2}, \tan\frac{\theta_1}{2} = \frac{\cos\alpha}{1-\sin\alpha}\tan\frac{\theta_4}{2}. \quad (5b)$$

Accordingly, $\theta_1$ and $\theta_3$ have opposite signs while $\theta_2$ and $\theta_4$ have the same for Eq. (5a). Eq. (5b) is on the contrary where $\theta_1$ and $\theta_3$ are of the same signs while $\theta_2$ and $\theta_4$ are of opposite. It means for each vertex, there are two types of M-V assignments. Therefore, there are totally 16 ($2^4$) types of M-V assignments of the square-twist origami pattern. Removing the duplicated and mirror-imaged cases that have identical kinematics, there are four M-V assignments of the square-twist origami pattern, noted as Type 1, Type 2, Type 3 and Type 4 as shown in Fig. 2. Here, the mountain and valley creases are illustrated by solid and dashed lines, respectively.

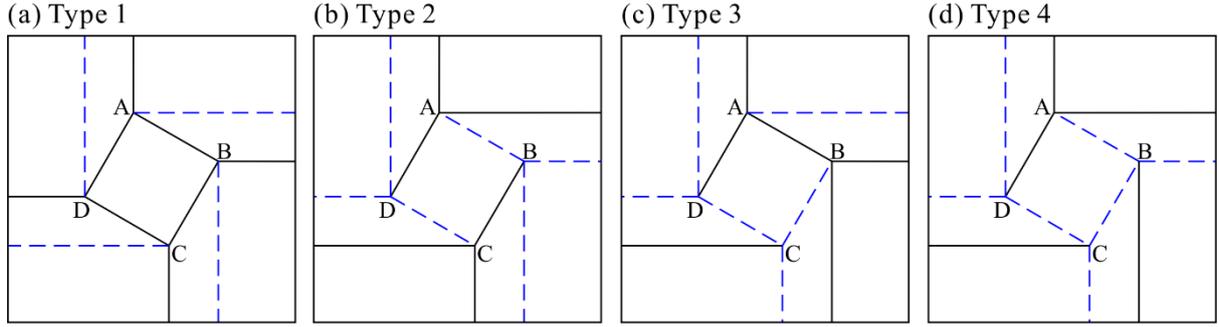

Fig. 2 Four types of square-twist origami patterns with different M-V assignments

To analyze the rigid foldability of these four types of square-twist origami patterns, the kinematic method based on the motion transmission path [23] is adopted. In the square-twist origami pattern, four 4-crease vertices A, B, C, and D form a closed loop of four spherical 4R linkages as shown in Fig. 1(c). If taking $\theta_1^a$ (rotation angle of the joint $a_1$) as the input of linkage A, we get the output of linkage A as $\theta_2^a$. Between linkages A and B, joints $a_2$ and $b_1$ are physically jointed together, so they have the same rotation, i.e., $\theta_2^a = \theta_1^b$. Then $\theta_1^b$ and $\theta_2^b$ are taken as the input and output of linkage B, respectively. Further, the motion is transmitted to linkages C and D with $\theta_2^b = \theta_1^c$ and $\theta_2^c = \theta_1^d$, where $\theta_2^d$ is the output of linkage D. To form the loop closure of this linkage network, the extra condition that $\theta_2^d = \theta_1^a$ must be satisfied. Therefore, the kinematic motion transmission path of the closed loop is

$$\theta_1^a \to \theta_2^a = \theta_1^b \to \theta_2^b = \theta_1^c \to \theta_2^c = \theta_1^d \to \theta_2^d$$
$$\theta_1^a = \theta_2^d \qquad (6)$$

Eq. (6) is the motion compatibility condition of the loop and it will be satisfied only when the origami pattern is rigidly foldable. Next, we will use this compatibility condition to check the rigid foldability of the four types of square-twist origami patterns.

## 2.1 Type 1

For the Type 1 square-twist origami pattern (Fig. 3), the M-V assignments of all the four vertices are identical, resulting in the same input-output relationship for each vertex. Since $\theta_1$ and $\theta_3$ have opposite signs, we have the kinematic relationship of each vertex as Eq. (5a) and

$$\tan\frac{\theta_2^d}{2} = \frac{\cos^4\alpha}{(1-\sin\alpha)^4}\tan\frac{\theta_1^a}{2}. \tag{7}$$

Since $\alpha \in (0, \pi/2)$, $\theta_2^d \neq \theta_1^a$, which means the compatibility condition is not satisfied, i.e., the Type 1 pattern is non-rigidly foldable. This can also be manifested in the kinematic motion transmission path represented by the arrowed line as shown in Fig. 3, where $\theta_2^d$ and $\theta_1^a$ cannot be connected by the curve $\theta_2^d = \theta_1^a$, i.e., $\theta_2^d \neq \theta_1^a$.

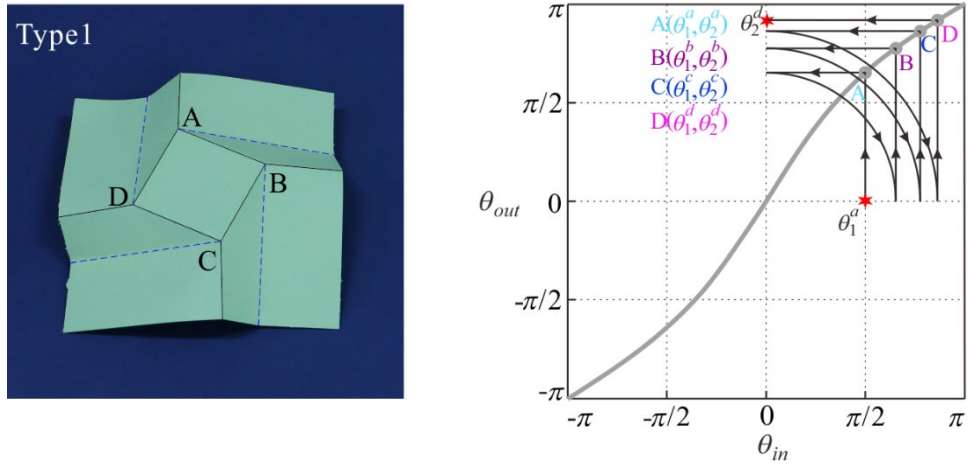

Fig. 3 Prototype of the Type 1 square-twist origami pattern and its corresponding kinematic motion transmission path. The kinematic curve is plotted when $\alpha = \pi/6$.

## 2.2 Type 2

For the Type 2 square-twist origami pattern (Fig. 4), the M-V assignments of alternative vertices, i.e., vertices A and C, vertices B and D, are identical. Actually, the M-V assignment of vertex A is the inverted configuration of that of vertex B, so they have identical input-output relationship. Since $\theta_1$ and $\theta_3$ have the same signs, we have the kinematic relationship of each vertex as Eq. (5b) and accordingly

$$\tan\frac{\theta_2^d}{2} = \frac{\cos^4\alpha}{(1+\sin\alpha)^4}\tan\frac{\theta_1^a}{2}. \tag{8}$$

Since $\alpha \in (0, \pi/2)$, $\theta_2^d \neq \theta_1^a$, the compatibility condition is neither satisfied in this case, i.e., the Type 2 pattern is also non-rigidly foldable. The kinematic motion transmission path is plotted in Fig. 4, where $\theta_2^d \neq \theta_1^a$ holds as well.

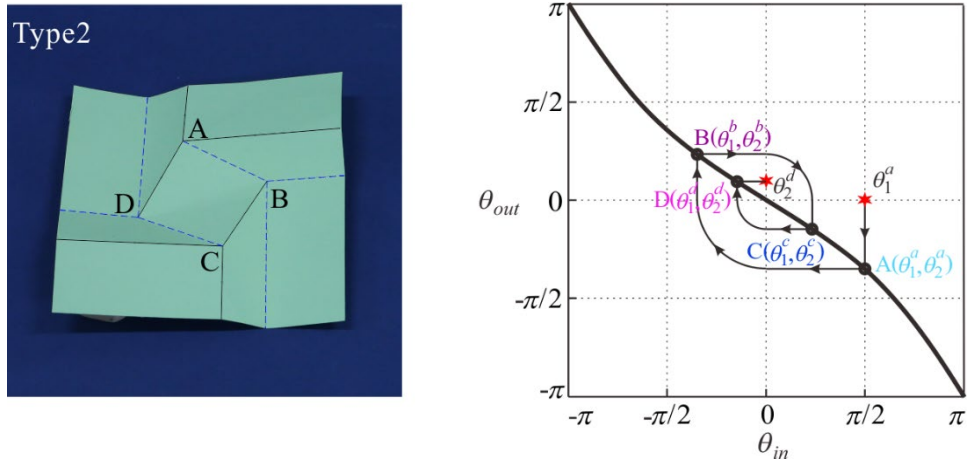

Fig. 4 Prototype of the Type 2 square-twist origami pattern and its corresponding kinematic motion transmission path. The kinematic curve is plotted when $\alpha = \pi/6$.

## 2.3 Type 3

In the Type 3 square-twist origami pattern (Fig. 5), the M-V assignment of each vertex is different. However, only two unique types exist. The M-V assignment of vertex A and that of vertex C are inverted, so they have identical input-output relationship as Eq. (5a) where $\theta_1$ and $\theta_3$ have opposite signs. The same case holds for the alternative vertices B and D with identical signs of $\theta_1$ and $\theta_3$, and the input-output relationship is given in Eq. (5b). Plotting the kinematic motion transmission path as shown in Fig. 5, we can find that $\theta_2^d$ and $\theta_1^a$ is connected by the curve $\theta_2^d = \theta_1^a$, indicating the satisfaction of the motion compatibility condition. Therefore, the Type 3 pattern is rigidly foldable, and the kinematic equations of the whole structure are

$$\tan\frac{\theta_2^a}{2} = \frac{\cos\alpha}{1-\sin\alpha}\tan\frac{\theta_1^a}{2}, \theta_3^a = -\theta_1^a, \theta_4^a = \theta_2^a; \tag{9a}$$

$$\theta_1^b = \theta_2^a, \tan\frac{\theta_2^b}{2} = \frac{-\cos\alpha}{1+\sin\alpha}\tan\frac{\theta_1^b}{2}, \theta_3^b = \theta_1^b, \theta_4^b = -\theta_2^b; \tag{9b}$$

$$\theta_1^c = \theta_2^b, \tan\frac{\theta_2^c}{2} = \frac{\cos\alpha}{1-\sin\alpha}\tan\frac{\theta_1^c}{2}, \theta_3^c = -\theta_1^c, \theta_4^c = \theta_2^c; \tag{9c}$$

$$\theta_1^d = \theta_2^c,\ \theta_2^d = \theta_1^a,\ \theta_3^d = \theta_1^d,\ \theta_4^d = -\theta_2^d. \tag{9d}$$

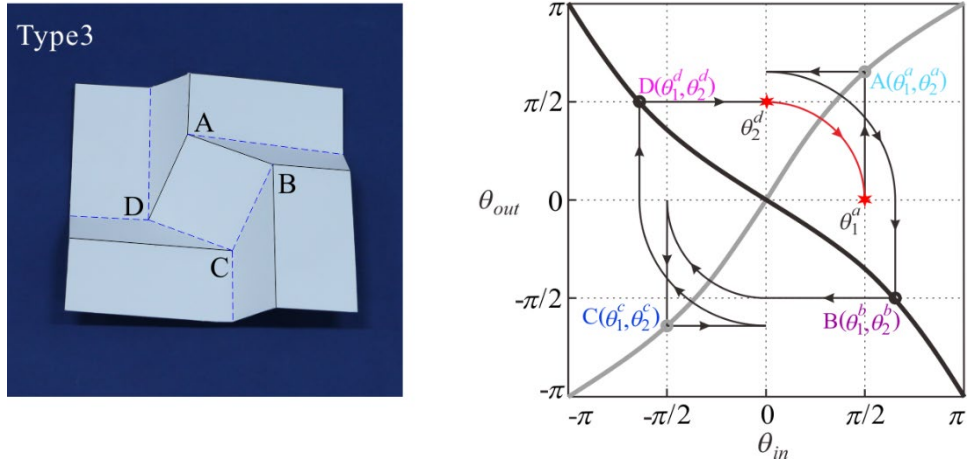

Fig. 5 Prototype of the Type 3 square-twist origami pattern and its corresponding kinematic motion transmission path. The kinematic curve is plotted when $\alpha = \pi/6$.

## 2.4 Type 4

In the Type 4 square-twist origami pattern (Fig. 6), there are three types of M-V assignments for all vertices, where vertices B and C have the same assignments. The M-V assignment of vertex A is the duplicated case of vertex D, so only two distinct types exist. In vertex A, $\theta_1$ and $\theta_3$ have the same signs, and the input-output relationship is Eq. (5b), while in vertex B, $\theta_1$ and $\theta_3$ have opposite signs, and the input-output relationship is given in Eq. (5a). The kinematic motion transmission path is plotted in Fig. 6, where we can find $\theta_2^d = \theta_1^a$, which means the motion compatibility condition is fulfilled in this case. Therefore, the Type 4 pattern is rigidly foldable, and its kinematic equations are

$$\tan\frac{\theta_2^a}{2} = \frac{-\cos\alpha}{1+\sin\alpha}\tan\frac{\theta_1^a}{2},\ \theta_3^a = -\theta_1^a,\ \theta_4^a = \theta_2^a; \tag{10a}$$

$$\theta_1^b = \theta_2^a,\ \tan\frac{\theta_2^b}{2} = \frac{\cos\alpha}{1-\sin\alpha}\tan\frac{\theta_1^b}{2},\ \theta_3^b = \theta_1^b,\ \theta_4^b = -\theta_2^b; \tag{10b}$$

$$\theta_1^c = \theta_2^b,\ \tan\frac{\theta_2^c}{2} = \frac{\cos\alpha}{1-\sin\alpha}\tan\frac{\theta_1^c}{2},\ \theta_3^c = -\theta_1^c,\ \theta_4^c = \theta_2^c; \tag{10c}$$

$$\theta_1^d = \theta_2^c, \theta_2^d = \theta_1^a, \theta_3^d = \theta_1^d, \theta_4^d = -\theta_2^d. \tag{10d}$$

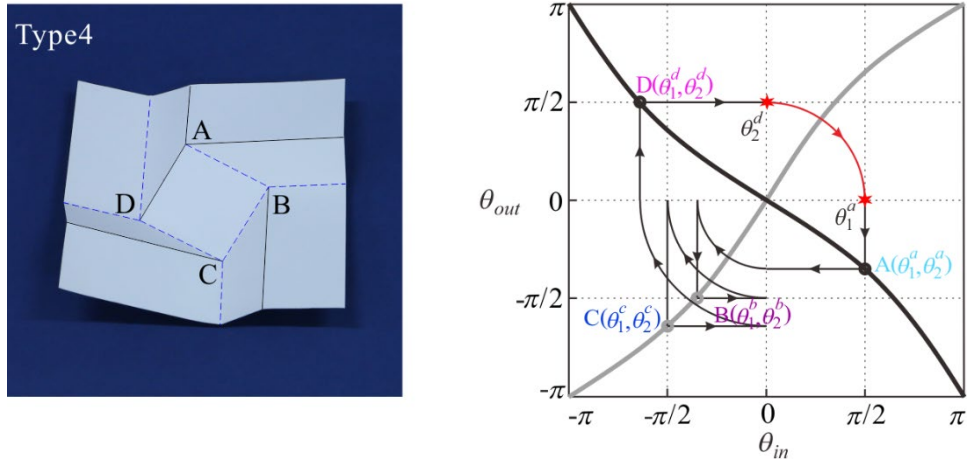

Fig. 6 Prototype of the Type 4 square-twist origami pattern and its corresponding kinematic motion transmission path. The kinematic curve is plotted when $\alpha = \pi/6$.

In summary, there are four distinct M-V assignments of the square-twist origami pattern. Type 1 and Type 2 are non-rigidly foldable, while Type 3 and Type 4 are rigidly foldable. We can choose the proper type according to our requirement on its rigid foldability. In the following, we are going to seek a way to convert the non-rigid case into a rigid one.

## 3. Conversion of rigid foldability

The rigid foldability analysis of the four square-twist patterns implies that we can convert their foldability from non-rigid to rigid by changing the relationship between the kinematic variables $\theta_1$ and $\theta_2$ of the vertices in the linkage loop. To change the input-output relationship of the non-rigid patterns, we add a diagonal crease in the central twisted square of the Type 1 and Type 2 pattern to constitute its corresponding modified pattern that consists of two 4-crease vertices and two 5-crease vertices as shown in Fig. 7. The 5-crease vertex can be modeled as a spherical 5R linkage in the kinematic analysis, and the modified pattern is modeled as a closed-loop network of spherical 4R and 5R linkages.

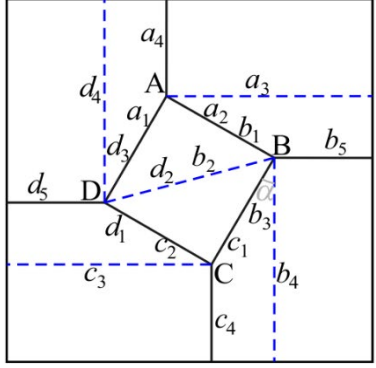
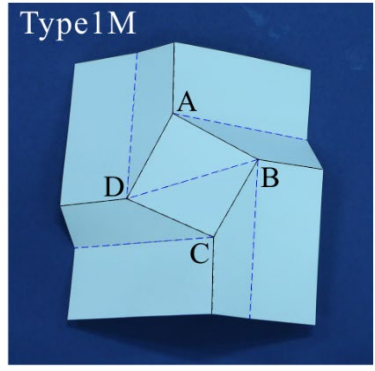
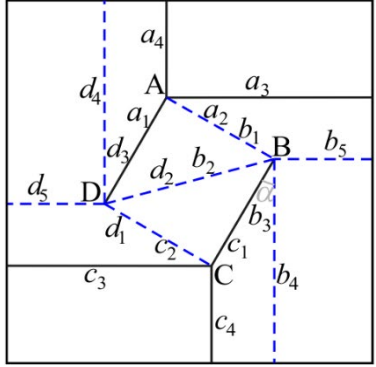
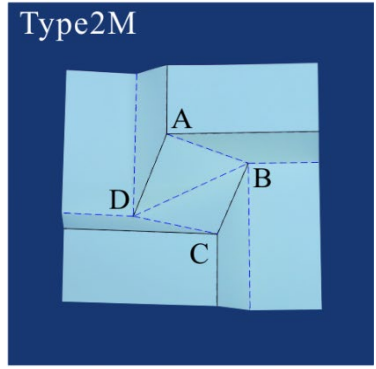

Fig. 7 The M-V assignments and paper prototypes of modified square-twist origami patterns: (a) modified Type 1, and (b) modified Type 2.

For the 5-crease vertex in the modified patterns, the geometrical parameters of the equivalent spherical 5$R$ linkage are defined as

$$\alpha_{12} = \frac{\pi}{4}, \alpha_{23} = \frac{\pi}{4}, \alpha_{34} = \alpha, \alpha_{45} = \frac{\pi}{2}, \alpha_{51} = \pi - \alpha. \tag{11}$$

Using the loop closure equation as in Eq. (1), the relationship between the kinematic variables $\theta_i$ and $\theta_{i+1}$ ($i$=1, 2, 3, 4, 5) can be obtained and simplified as

$$\tan\frac{\theta_4}{2} = \frac{Q_2 \pm \sqrt{Q_3}}{Q_1}$$
$$Q_1 = \sin\alpha + \cos\alpha + \sin\alpha \cdot \cos\theta_1 + \cos\alpha \cdot \cos\theta_3$$
$$Q_2 = -\sin\theta_3 \tag{12a}$$
$$Q_3 = \sin^2\alpha \cdot (\sin^2\theta_1 + \sin^2\theta_3) - 2\sin\alpha \cdot \cos\alpha \cdot (\cos\theta_1 - \cos\theta_3)$$

$$\tan\frac{\theta_5}{2} = \frac{-P_2 \pm \sqrt{P_2^2 - P_1 \cdot P_3}}{P_1}$$

$$\begin{aligned}
P_1 &= (\cos\alpha \cdot \sin\theta_3 \cdot \sin\theta_4 - \cos^2\alpha \cdot \cos\theta_3 \cdot \cos\theta_4 - \sin\alpha \cdot \cos\alpha \cdot \cos\theta_4 + 1) \\
&\quad + (\sin^2\alpha \cdot \cos\theta_3 - \sin\alpha \cdot \cos\alpha) \\
P_2 &= -(\sin\alpha \cdot \sin\theta_3 \cdot \cos\theta_4 + \sin\alpha \cdot \cos\alpha \cdot \cos\theta_3 \cdot \sin\theta_4 + \sin^2\alpha \cdot \sin\theta_4) \\
P_3 &= (\cos\alpha \cdot \sin\theta_3 \cdot \sin\theta_4 - \cos^2\alpha \cdot \cos\theta_3 \cdot \cos\theta_4 - \sin\alpha \cdot \cos\alpha \cdot \cos\theta_4 + 1) \\
&\quad - (\sin^2\alpha \cdot \cos\theta_3 - \sin\alpha \cdot \cos\alpha)
\end{aligned} \qquad (12b)$$

$$\cos\theta_2 = 2\sin\alpha \cdot (-\cos\alpha \cdot \cos\theta_4 + \cos\alpha \cdot \cos\theta_5 - \sin\alpha \cdot \sin\theta_4 \cdot \sin\theta_5) + 1. \qquad (12c)$$

With this relationship, we are able to conduct the kinematic and rigid foldability analysis of the modified Type 1 and Type 2 square-twist origami patterns, noted as Type 1M and Type 2M.

To check the rigid foldability of modified square-twist patterns consisting of both 4-crease and 5-crease vertices, one more condition should be satisfied as follows besides the compatibility condition in Eq. (6).

$$\varphi_2^d = \varphi_2^b \qquad (13)$$

Here, $\varphi_2^b$ and $\varphi_2^d$ are the dihedral angles between the two facets with a common crease $b_2$ and $d_2$, respectively. The dihedral angles of the 5-crease vertices B and D can be calculated by the kinematics of the spherical 5R linkage with two input angles being $\varphi_1$ and $\varphi_3$. In the linkage loop of the modified pattern, the two input angles are transmitted from the 4-crease vertices A and C. The transmission relationship between 4- and 5-crease vertices are $\varphi_2^a = \varphi_1^b$, $\varphi_2^c = \varphi_1^d$, $\varphi_1^a = \varphi_3^d$, and $\varphi_1^c = \varphi_3^b$. To meet the extra compatibility condition in Eq. (13), vertices B and D that have identical geometrical parameters should have the same kinematic relationships between dihedral angles. So we have $\varphi_4^d = \varphi_4^b$, $\varphi_5^d = \varphi_5^b$, $\varphi_1^d = \varphi_1^b$ and $\varphi_3^d = \varphi_3^b$. Accordingly, $\varphi_2^a = \varphi_2^c$ and $\varphi_1^a = \varphi_1^c$. The kinematic relationships of vertices A and C are identical as well, resulting in $\varphi_3^a = \varphi_3^c$ and $\varphi_4^a = \varphi_4^c$. Thus, the relationship between the dihedral angles of the modified patterns can be expressed as

$$\begin{aligned}
&\varphi_1^b = \varphi_1^d = \varphi_2^c = \varphi_4^c = \varphi_2^a = \varphi_4^a, \varphi_3^b = \varphi_3^d = \varphi_1^a = \varphi_3^a = \varphi_1^c = \varphi_3^c, \\
&\varphi_2^b = \varphi_2^d, \varphi_4^b = \varphi_4^d, \varphi_5^b = \varphi_5^d.
\end{aligned} \qquad (14)$$

It can be found from Eq. (14) that the compatibility conditions in Eqs. (6) and (13) are both satisfied, which means the modified square-twist patterns in Fig. 7 are rigidly foldable. Next, we will investigate their kinematics and bifurcation behaviour.

## 4. Kinematics of modified square-twist patterns and their bifurcation behaviour

### 4.1 Type 1M

In the 5-crease vertices of the Type 1M pattern in Fig. 7(a), the relationship between kinematic variables and the dihedral angles is

$$\theta_1 = \pi - \varphi_1,\ \theta_2 = \varphi_2 - \pi,\ \theta_3 = \pi - \varphi_3,\ \theta_4 = \varphi_4 - \pi,\ \theta_5 = \pi - \varphi_5. \tag{15}$$

Substituting Eq. (15) to Eq. (12), the relationship between the dihedral angles of the 5-crease vertex can be expressed as

$$\begin{aligned}
&\tan\frac{\varphi_4}{2} = \frac{q_1}{q_2 \pm \sqrt{q_3}} \\
&q_1 = \sin\alpha + \cos\alpha - \sin\alpha \cdot \cos\varphi_1 - \cos\alpha \cdot \cos\varphi_3 \\
&q_2 = \sin\varphi_3 \\
&q_3 = \sin^2\alpha \cdot (\sin^2\varphi_1 + \sin^2\varphi_3) + 2\sin\alpha \cdot \cos\alpha \cdot (\cos\varphi_1 - \cos\varphi_3)
\end{aligned}, \tag{16a}$$

$$\begin{aligned}
&\tan\frac{\varphi_5}{2} = \frac{p_1}{p_2 + \sqrt{p_2^2 - p_1 \cdot p_3}} \\
&p_1 = (\cos\alpha \cdot \sin\varphi_3 \cdot \sin\varphi_4 + \cos^2\alpha \cdot \cos\varphi_3 \cdot \cos\varphi_4 - \sin\alpha \cdot \cos\alpha \cdot \cos\varphi_4 - 1) \\
&\quad + (\sin^2\alpha \cdot \cos\varphi_3 + \sin\alpha \cdot \cos\alpha) \\
&p_2 = \sin\alpha \cdot \sin\varphi_3 \cdot \cos\varphi_4 - \sin\alpha \cdot \cos\alpha \cdot \cos\varphi_3 \cdot \sin\varphi_4 + \sin^2\alpha \cdot \sin\varphi_4 \\
&p_3 = (\cos\alpha \cdot \sin\varphi_3 \cdot \sin\varphi_4 + \cos^2\alpha \cdot \cos\varphi_3 \cdot \cos\varphi_4 - \sin\alpha \cdot \cos\alpha \cdot \cos\varphi_4 - 1) \\
&\quad - (\sin^2\alpha \cdot \cos\varphi_3 + \sin\alpha \cdot \cos\alpha)
\end{aligned}, \tag{16b}$$

$$\cos\varphi_2 = 2\sin\alpha \cdot (-\cos\alpha \cdot \cos\varphi_4 + \cos\alpha \cdot \cos\varphi_5 - \sin\alpha \cdot \sin\varphi_4 \cdot \sin\varphi_5) - 1. \tag{16c}$$

For the 4-crease vertices, regarding the relationship between the kinematic variables and the dihedral angles, we have

$$\theta_1 = \pi - \varphi_1,\ \theta_2 = \pi - \varphi_2,\ \theta_3 = \varphi_3 - \pi,\ \theta_4 = \pi - \varphi_4. \tag{17}$$

Accordingly, we can obtain the relationship between the dihedral angles of vertices A and C by simplifying Eq. (5a),

$$\tan\frac{\varphi_2}{2} = \frac{\cos\alpha}{1+\sin\alpha}\tan\frac{\varphi_1}{2},\ \varphi_3 = \varphi_1,\ \varphi_4 = \varphi_2. \tag{18}$$

Therefore, Eqs. (14), (16) and (18) form the whole set of kinematic relationship between the dihedral angles of the Type 1M pattern. From Eq. (16a), we can find that there are two solutions for $\varphi_4^b$ and $\varphi_4^d$, indicating the existence of two kinematic paths for vertices B and D in the Type 1M pattern. Figure 8(a) presents the relationship between the dihedral angles of vertex B in two kinematic paths when the twist angle $\alpha = \pi/6$. Two phenomena should be noticed here. First, for Path 1, $\varphi_5^b < 0$ between the configuration I ($\varphi_1^b = 0$) and II ($\varphi_1^b = 0.41\pi$), meaning that there is facet interference between these two configurations. For Path 2, $\varphi_4^b > \pi$ between the configuration III ($\varphi_1^b = 0.41\pi$) and IV ($\varphi_1^b = \pi$), which indicates the crease 4 switch from the mountain crease to the valley one in this region. Redefining $\varphi_4^b$ as the smaller angle formed by two adjacent facets, the kinematic curve between $\varphi_4^b$ and $\varphi_1^b$ for Path 2 is replotted in the grey line in Fig. 8(b), where the configurations of the Type 1M pattern in the two paths are presented. Two pairs of parallel rectangle panels are distinctively colored to demonstrate the different configurations along two paths. The pattern only has bifurcation at the fully folded and fully deployed configuration. Once the motion is underway, the two kinematic paths cannot be switched each other. The configuration at $\varphi_1^b = \pi/4$ in Path 1 is physically impossible because of the interference of the facets, which corresponds to the case with a negative value of $\varphi_5^b$ in Fig. 8(a).

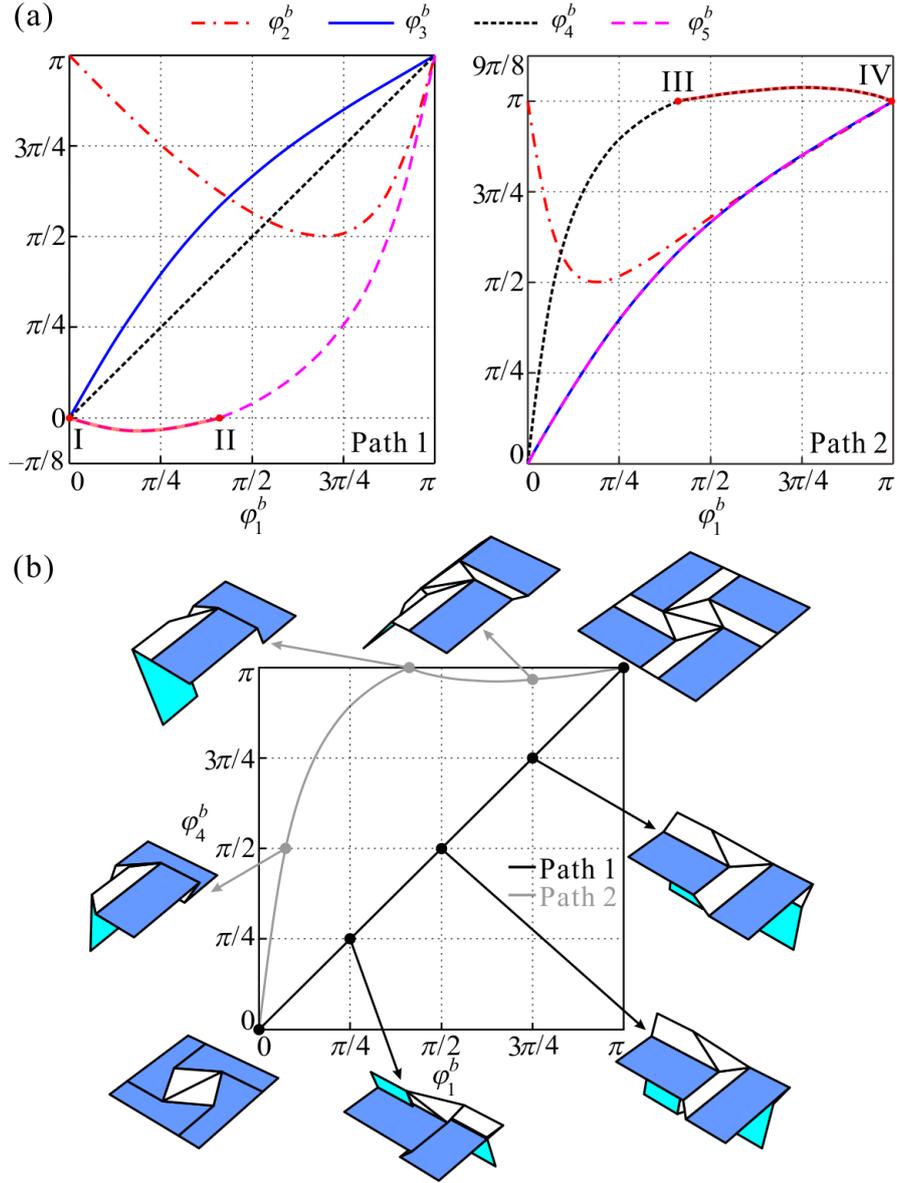

Fig. 8 Kinematic curves and bifurcation of the Type 1M pattern with $\alpha = \pi/6$: (a) kinematical curves of vertex B in the pattern, and (b) motion sequences of the pattern under two kinematic paths.

It is found that the configurations II and III in Fig. 8 (a) have the same value of $\varphi_1^b$, but this relationship only holds when $\alpha = \pi/6$. To note, the facet interference and switch between mountain and valley creases are both related to $\alpha$. Figure 9 presents the variation in kinematic curves for the two paths when $\alpha$ is varied. The facet interference appears when $\alpha \geq 23.4°$ and the range of motion where interference happens increase with $\alpha$. The switch of mountain and valley crease also emerges with the increase of $\alpha$. The critical point locates at the same position where $\alpha = 23.4°$. The larger $\alpha$ is, the sooner the crease switch happens.

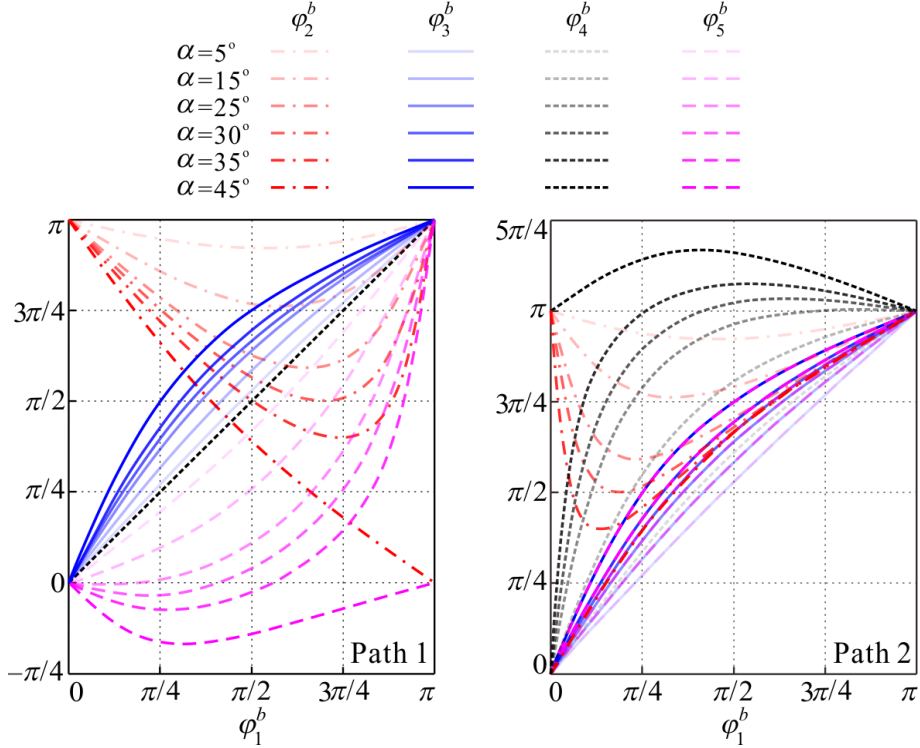

Fig. 9 Kinematic curves of the Type 1M pattern with varied $\alpha$

## 4.2 Type 2M

Similarly, we can also analyze the kinematics of the Type 2M pattern in Fig. 7(b) by regarding it as a closed-loop network consisting of spherical 4R and 5R linkages. In the 5-crease vertex of the Type 2M pattern, the relationship between kinematic variables and the dihedral angles is

$$\theta_1 = \varphi_1 - \pi, \theta_2 = \varphi_2 - \pi, \theta_3 = \pi - \varphi_3, \theta_4 = \varphi_4 - \pi, \theta_5 = \varphi_5 - \pi. \tag{19}$$

Substituting Eq. (19) to Eq. (12), the relationship between the dihedral angles of the 5-crease vertices B and D can be expressed as

$$\tan\frac{\varphi_4}{2} = \frac{q_1}{q_2 \pm \sqrt{q_3}}$$
$$q_1 = \sin\alpha + \cos\alpha - \sin\alpha \cdot \cos\varphi_1 - \cos\alpha \cdot \cos\varphi_3$$
$$q_2 = \sin\varphi_3 \tag{20a}$$
$$q_3 = \sin^2\alpha \cdot (\sin^2\varphi_1 + \sin^2\varphi_3) + 2\sin\alpha \cdot \cos\alpha \cdot (\cos\varphi_1 - \cos\varphi_3)$$

$$\tan\frac{\varphi_5}{2} = \frac{p_1}{-p_2 + \sqrt{p_2^2 - p_1 \cdot p_3}}$$

$$p_1 = (\cos\alpha \cdot \sin\varphi_3 \cdot \sin\varphi_4 + \cos^2\alpha \cdot \cos\varphi_3 \cdot \cos\varphi_4 - \sin\alpha \cdot \cos\alpha \cdot \cos\varphi_4 - 1)$$
$$+ (\sin^2\alpha \cdot \cos\varphi_3 + \sin\alpha \cdot \cos\alpha) \quad , \quad (20b)$$

$$p_2 = \sin\alpha \cdot \sin\varphi_3 \cdot \cos\varphi_4 - \sin\alpha \cdot \cos\alpha \cdot \cos\varphi_3 \cdot \sin\varphi_4 + \sin^2\alpha \cdot \sin\varphi_4$$

$$p_3 = (\cos\alpha \cdot \sin\varphi_3 \cdot \sin\varphi_4 + \cos^2\alpha \cdot \cos\varphi_3 \cdot \cos\varphi_4 - \sin\alpha \cdot \cos\alpha \cdot \cos\varphi_4 - 1)$$
$$- (\sin^2\alpha \cdot \cos\varphi_3 + \sin\alpha \cdot \cos\alpha)$$

$$\cos\varphi_2 = 2\sin\alpha \cdot (-\cos\alpha \cdot \cos\varphi_4 + \cos\alpha \cdot \cos\varphi_5 + \sin\alpha \cdot \sin\varphi_4 \cdot \sin\varphi_5) - 1. \quad (20c)$$

In the 4-crease vertices A and C of Type 2M pattern, the relationship between the kinematic variables and the dihedral angles is

$$\theta_1 = \pi - \varphi_1, \; \theta_2 = \varphi_2 - \pi, \; \theta_3 = \pi - \varphi_3, \; \theta_4 = \pi - \varphi_4. \quad (21)$$

Substituting Eq. (21) into Eq. (5b), we have the relationship between the dihedral angles of vertices A and C as

$$\tan\frac{\varphi_2}{2} = \frac{\cos\alpha}{1 - \sin\alpha} \tan\frac{\varphi_1}{2}, \; \varphi_3 = \varphi_1, \; \varphi_4 = \varphi_2. \quad (22)$$

Therefore, Eqs. (14), (20) and (22) form the whole set of kinematic relationship between the dihedral angles of the Type 2M pattern. The kinematic curves of this pattern with $\alpha = \pi/6$ are plotted in Fig. 10, where two kinematic paths exist, indicating the bifurcation of this pattern. The kinematic curves of Path 1 and Path 2 are presented as solid and dashed lines, respectively. Similar to the Type 1M pattern, bifurcation would happen at the fully folded configuration I and the fully deployed configuration III. One more bifurcation occurs at the configuration II during the motion process where $\varphi_1^b = 0.59\pi$. The Type 2M pattern can move smoothly along either kinematic path, and bifurcate to the other at one of these bifurcation configurations.

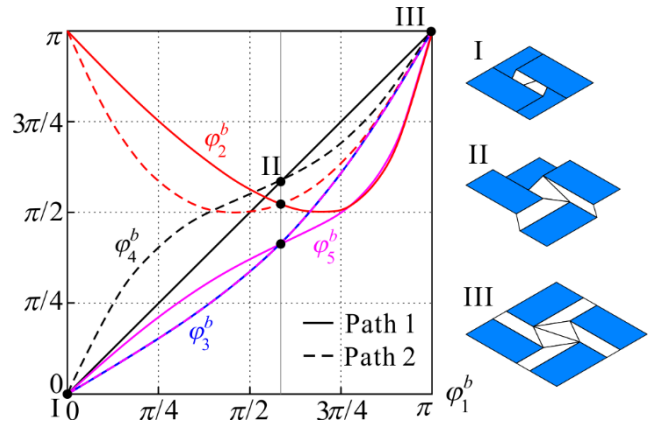

Fig. 10 Kinematic curves and bifurcation configurations of the Type 2M pattern with $\alpha = \pi/6$, where bifurcation would occur at configurations I, II and III.

## 5. Conclusions and discussion

In this paper, we have analyzed the rigid foldability and kinematics of the square-twist origami patterns with four types of M-V assignments. Their rigid foldability is analyzed by the kinematic method based on the motion transmission path. It has been found that two of the four patterns are rigidly foldable, while the other two are not. Using the equivalence between the rigid origami pattern and the closed-loop network of spherical linkages, the kinematic equations of the rigid cases have been derived. We have also proposed a method to convert the non-rigid square-twist origami patterns to rigid ones by adding a diagonal crease into the central twist square in the pattern, and have validated the rigid foldability of the modified patterns. The kinematics and motion behaviour of the modified square-twist patterns have been illustrated with explicit kinematic equations. Bifurcation phenomenon has been revealed in the two modified patterns, with facet interference checked and its relevance to the geometrical parameter sought. Both the two modified patterns have bifurcation at the fully folded or fully deployed configuration. The modified patterns will serve as new units to construct rigidly foldable origami tessellations.

This work not only helps us to better understand the motion of square-twist origami patterns and their variations, but also provides us a way to convert the rigid foldability of an origami pattern by adding creases. The placing of the extra creases needs further study. For example, instead of setting the additional crease in the central twist square in the square-twist pattern, we can also add the crease in any of the surrounding quadrangles. This approach of converting rigid foldability can be extended to other origami patterns to create more rigidly foldable patterns.

**Acknowledgements**

Y. C. acknowledges the support of the National Natural Science Foundation of China (Projects 51825503, 51721003), J. M. acknowledges the support of the National Natural Science Foundation of China (Projects 51575377).